\documentclass[conference]{IEEEtran}
\usepackage{cite}
\usepackage{amsmath,amssymb,amsfonts}
\usepackage{algorithmic}
\usepackage{graphicx}
\usepackage{textcomp}
\usepackage{xcolor}
\usepackage{algorithmic}
\usepackage{epigraph}
\usepackage{subfigure}
\usepackage{algorithm}

\def\BibTeX{{\rm B\kern-.05em{\sc i\kern-.025em b}\kern-.08em
    T\kern-.1667em\lower.7ex\hbox{E}\kern-.125emX}}
\begin{document}

\title{Attentive Geo-Social Group Recommendation}

\author{\IEEEauthorblockN{\small  1\textsuperscript{st}  Fei Yu}
\IEEEauthorblockA{\footnotesize \textit{School of Computer Science and Technology}\\
 \textit{Harbin Institute of Technology},\\ Harbin, 150001, China \\
yf@hit.edu.cn}
\and
\IEEEauthorblockN{\small 2\textsuperscript{nd} Feiyi Fan}
\IEEEauthorblockA{\footnotesize \textit{School of Computing} \\
\textit{National University of Singapore}\\
Singapore\\
fanfy@comp.nus.edu.sg}
\and
\IEEEauthorblockN{\small 3\textsuperscript{rd} Shouxu Jiang}
\IEEEauthorblockA{\footnotesize \textit{School of Computer Science and Technology}\\
\textit{Harbin Institute of Technology}, \\Harbin, 150001, China \\
jsx@hit.edu.cn}
\and
\IEEEauthorblockN{\small 4\textsuperscript{th} Kaiping Zheng}
\IEEEauthorblockA{\footnotesize \textit{School of Computing} \\
\textit{National University of Singapore}\\
Singapore\\
kaiping@comp.nus.edu.sg}
}

\maketitle

\begin{abstract}

Social activities play an important role in people's daily life since they interact. 
For recommendations based on social activities, it is vital to have not only the activity information but also individuals' social relations. 
Thanks to the geo-social networks and widespread use of location-aware mobile devices, massive geo-social data is now readily available for exploitation by the recommendation system.

In this paper, a novel group recommendation method, called attentive geo-social group recommendation, is proposed to recommend the target user with both activity locations and a group of users that may join the activities. 
We present an attention mechanism to model the influence of the target user $u_T$ in candidate user groups that satisfy the social constraints. It helps to retrieve the optimal user group and activity topic candidates, as well as explains the group decision-making process. Once the user group and topics are retrieved, a novel efficient spatial query algorithm SPA-DF is employed to determine the activity location under the constraints of the given user group and activity topic candidates.

The proposed method is evaluated in real-world datasets and the experimental results show that the proposed model significantly outperforms baseline methods.
\end{abstract}


\section{Introduction}
``Man is by nature a social animal,'' a quote made by Aristotle, the Greek philosopher, highlights the importance of social relations. The advancement of technology makes it easier for people to connect with each other via applications such as geo-social network (GeoSN).
Apart from the connections between people, GeoSNs enable users to post and share their visited physical locations or geo-tag information via ``check-in'', which establishes the link between the real world (offline) and social networks (online)\cite{Doytsher}.

Unfortunately, the existing group recommendation methods do not fully exploit the potential of massive geo-social data collected in the GeoSNs. Specifically, GeoSNs record heterogeneous data consisting of relations in society, geography, and geo-social space, while existing group recommendation methods \cite{Tran}\cite{Cao}\cite{YIN} focus on recommending single-typed content only, not meeting the increasing demand on recommendation quality and user experience.

\textbf{Motivation Example:}
We shall use a real-world scenario to back up the argument mentioned above. Alice loves tourism and is planning a trip during the public holiday. She has plenty of destination options such as beach resorts, historical sites, or national park self-driving tour. 
For the existing location recommendation methods, Alice's preference for location is taken into consideration, and the system may answer ``Mauritius'', ``Beijing, China'', and ``Yellowstone National Park, United States'' for her query. 

However, as social animals, most people prefer a trip with friends or family to travelling alone. 
Recommending only locations does not help much because Alice still has to check with her friends for each option to see if they are willing to travel with her. 
In such a case, she needs a recommendation on both the place to go and the friends who may accept her invitation. An ideal recommendation responding to her query might be ``Mauritius, Bob and Christina'', where Bob and Christina are two friends that also favour island tourism.

To tackle the recommendation problem mentioned above, we propose a novel type of group recommendation method that recommends both activities and a group of ``promising''\footnote{Users who are likely to join the recommended activity} users using geo-social data (i.e., data captured in GeoSNs), which is termed as \textbf{attentive geo-social group recommendation (AGSGR)}.

\begin{figure}[!t]
\centering
\includegraphics[width=0.88\linewidth]{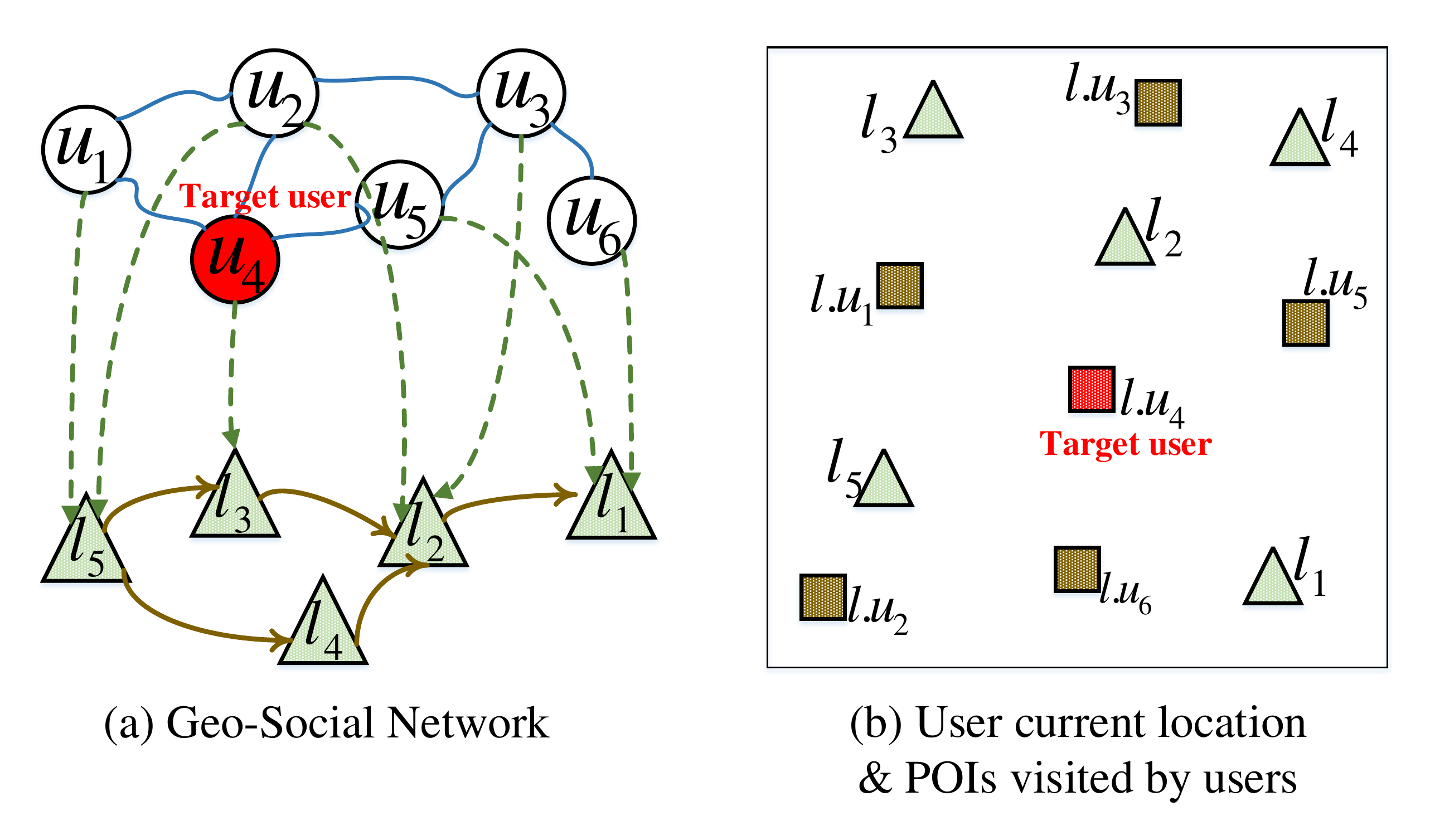}
\caption{An example of geo-social network.}
\label{fig:GSN}
\label{fig:framework}
\end{figure}

\textbf{Problem Statement:} 
As shown in Figure~\ref{fig:GSN}(a), a GeoSN can be represented by a heterogeneous graph, which consists of two types of nodes (user, location), three types of links (the social relationship, the association relationship between locations,  and the check-in link between users and activity locations). Figure~\ref{fig:GSN}(b) illustrates the users and point-of-interest (POI) in the geography space, where the rectangles represent users' current locations and the triangles denote the location of POI. 

Given the target user $u_4$ and an integer $h$, AGSGR returns an $h$-sized user group and top-$K$ activity locations which $u_4$ and the user group may be interested in. The recommendation results should satisfy the social constraint (i.e., \textbf{group cohesiveness}) and geography constraints (i.e., \textbf{geographical accessibility}). The multi-typed group recommendation problem is non-trivial because the recommendations made should balance the preferences of both the target user and the recommended user group as well as optimize the accessibility of recommended locations for all users in the group.

\textbf{Research Gaps:} Traditional group recommendation, geo-social query and event-partner recommendation are closely related to the aforementioned problem.

Traditional group recommendation methods recommend a group of users with items (users, activities, etc.) \cite{Tran}\cite{Cao}\cite{YIN}.
Given a query user $u_q$ and a constant $k$, geo-social query finds a $k$-sized set of users subject to pre-defined social constraints. The research gap of both traditional group recommendation and geo-social query is that they return single-typed content only.

Recently, event-partner recommendation is proposed as a new recommendation paradigm, including event-based partner recommendation, partner-based event recommendation, and joint event-partner recommendation.

Event-based partner recommendation helps the target user find a partner to participate in an activity together while partner-based event recommendation allows the target user and his partner to find activities to join in. 
They return single-typed recommendation results. Recommendation results of joint event-partner recommendations, such as the graph based embedding model (GEM) proposed in\cite{Hongzhi}, are multi-typed. Though it helps the target user find a partner and an event simultaneously, it recommends only one user with one event for the target user, which limits its application. 

\textbf{Solution:} To resolve the multi-typed group recommendation problem, we propose to exploit and simulate the decision-making process of group activity planning. 
We first select a user group by optimizing the target user's decision-making power (influence) on the other users subject to activity topics, then select top-$k$ activity locations by optimizing the recommended users' accessibility (travel distance) under the constraint of activity topics. For the former step, we propose to model the target users' group-wise decision-making power using an attention mechanism. 
For the latter step, an improved single point aggregation algorithm, SPA-DF, is proposed to prune the activity location search space; thus, the location query can be done with enhanced efficiency.

AGSGR can be seen as a generalized form of traditional group recommendation which recommends multi-typed content. It also extends the existing geo-social query by considering the decision-making process and the preferences of the users in the user group. Furthermore, the joint event-partner recommendation can be seen as a special case of our proposed AGSGR.

\textbf{Contributions:} 
We summarize the key contributions as follows:
\begin{itemize}
\item We formulate a novel type of group recommendation problem: multi-typed geo-social group recommendation. 
\item We develop a recommendation method by exploiting the attention mechanism to dynamically adjust the target user's influence across different user groups, explaining the decision-making process of group activity planning better.
\item We propose an efficient spatial search method SPA-DF to find activity locations under the constraints of the given user group and activity topics.
\item In comparison, we evaluate the effectiveness of our proposed method with the baseline methods on real-world datasets. 
The results show that the proposed method is more superior than the baseline methods in terms of recommendation accuracy.
\end{itemize}

The remainder of the paper is organized as follows. Section 2 formalizes problem-related definitions. Section 3 details the proposed AGSGR method. The experiment results are reported in Section 4. Related work is reviewed in Section 5, and conclusions are drawn in Section 6.

\newtheorem{definition}{Definition}

\section{Preliminaries}
In this section, definitions are introduced first and then relevant notations are summarized in Table \ref{tab0}.

\begin{definition}
\rm \textbf{Geo-Social Network (GeoSN):} A GeoSN $\mathcal{G}=$ $<\mathcal{V},\mathcal{E}>$ is a directed heterogeneous graph. Node set $\mathcal{V} = \{U, L\}$, where $U$ is the user node set and $L$ is the POI node set. Edge set $\mathcal{E}=\{\mathcal{E}_{u,u}, \mathcal{E}_{u,l},\mathcal{E}_{l,l}\}$, where $\mathcal{E}_{u,u}$ denotes the social links, $\mathcal{E}_{l,l}$ denotes the association relationship between POIs, and $\mathcal{E}_{u,l}$ denotes the check-in link between users and POIs.
\end{definition}

\begin{definition}
\rm \textbf{User Activity:} In GeoSNs, a user activity is a triplet $(u, x, \ell_x)$ which represents user $u$ selects the topic $x$-related service in the POI $l_x$, and checks in $\ell_x$. $\ell_x$ is the \textbf{activity location} where the user participates in the activity.
\end{definition}

As shown in Figure~\ref{fig:GSN}, the social network $\mathcal{G}(U)=<U,\mathcal{E}_{u,u}>$ is a subgraph of GeoSN $\mathcal{G}$, where $U \subseteq \mathcal{V}, \mathcal{E}_{u,u}\subseteq \mathcal{E}$.

The metric we employ to measure the social acquaintance level of a user group is the group cohesiveness, which can be quantified by \textbf{$k$-core} \cite{Seidman}.

\begin{definition}
\rm \textbf{$k$-core.} For a social network (SN) $\mathcal{G}(U)=<U,\mathcal{E}_{u,u}>$, which is a subgraph of GeoSN $\mathcal{G}$, the connected subgraph of $\mathcal{G}(U)$: $\mathcal{G}(U')=<U',\mathcal{E}_{u,u}^{'}>$ is a $k$-core if every user node $u\in U'$ has at least degree $k$, where $U' \subseteq U,\; \mathcal{E}_{u,u}^{'}  \subseteq \mathcal{E}_{u,u}$.
\end{definition}

\begin{figure}[!t]
\centering
\includegraphics[width=0.8\linewidth]{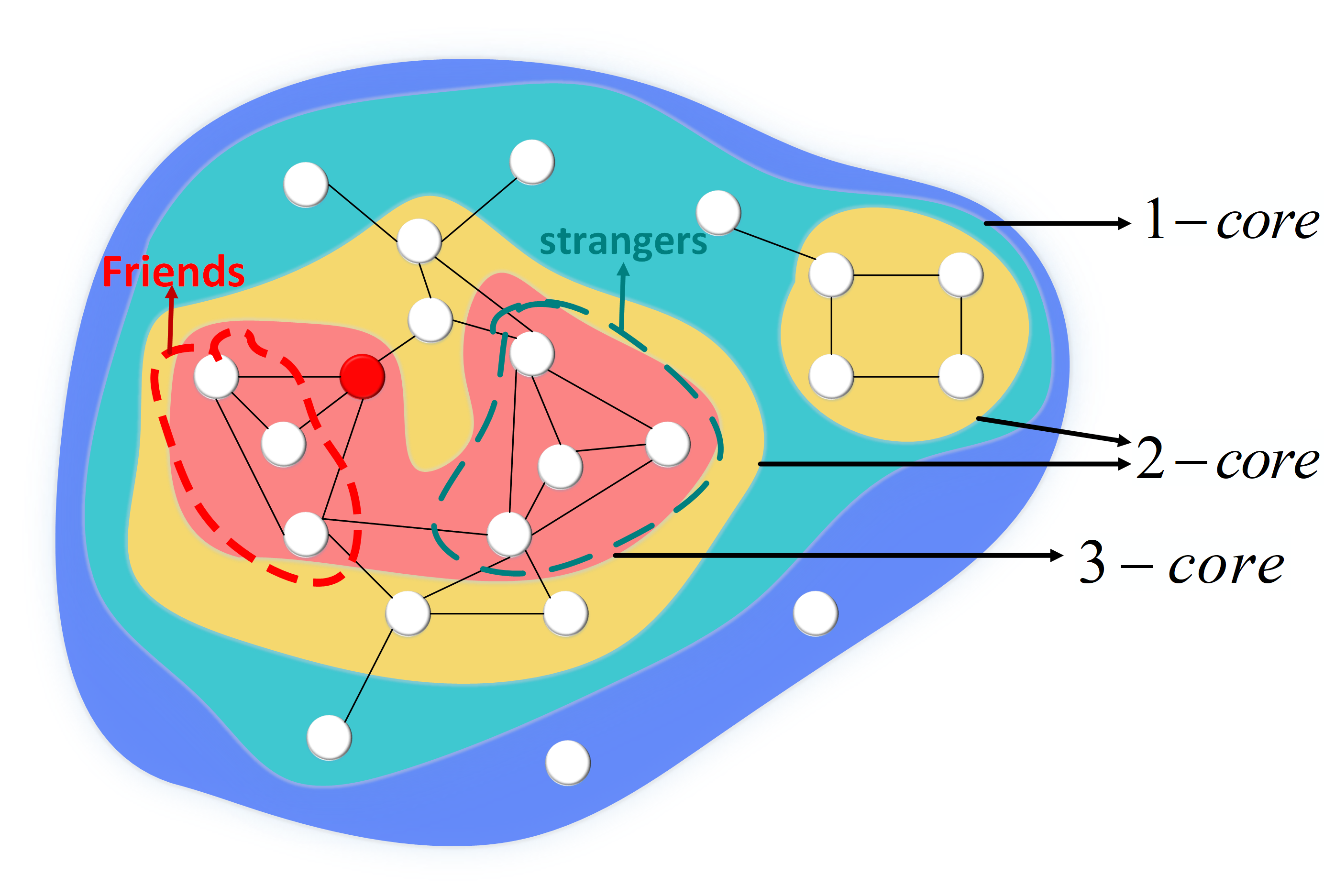}
\caption{ $k$-core decomposition for an example graph.} 
\label{fig:kcorefigure}
\end{figure}
It should be noted that a user group represented by a sub-graph satisfying the condition of $k$-core cannot guarantee that the target user has a relationship with every other user in the user group.

We take the GeoSN illustrated in Figure~\ref{fig:kcorefigure} as an example. The red circle represents the target user $u_T$. The recommendation system first selects the user group with $3$-core as the candidate user group. However, the user group with $3$-core contains only three friends of the target user $u_T$; the remainder of the user group are not friends of $u_T$.

\begin{definition}
\rm \textbf{AGSGR:} Given a social network $\mathcal{G}(U)$, a target user $u_T$, a AGSGR $Q=<u_T, h, k, \alpha>$ is a $4$-tuple, where $h$ is the size constraint of user group, $k$ represents the social acquaintance constraint $k$-core, $\alpha$ is 
the group evaluation index of the decision-making power, i.e. influence, of the target user $u_T$ in the user group. 
AGSGR returns a user set $U^{\star} \subseteq U$ and one activity location $\ell$ that:
\begin{itemize}
\item the subgraph $\mathcal{G}(U^{\star})$ is $k$-core and connected component;
\item the size of the user group $U^{\star}$ is $h$;
\item the decision-making power of $u_T$ in $U^{\star} \in \{U_{(j)}^k\}_{j=1}^n$ under its corresponding  topic $T^{\star}$, i.e.  $\alpha_{u_T}(U^{\star})$ is maximized. $\{U_{(j)}^k\}_{j=1}^n$ is the subgraph set with $k$-core of $\mathcal{G}(U)$; 
\item the activity location $\ell$ not only belongs to the corresponding activity topic $T^{\star}$, but also is closest to each user in the user group $U^{\star}$.
\end{itemize}
\end{definition}

\begin{table}[htbp]
\footnotesize
\caption{Symbol Description}
\label{tab0}
	\centering
	\begin{tabular}{c|c}
		\hline\hline
		\textbf{Symbol}&\textbf{Description} \\ 
		\hline
		$\mathcal{G}=<\mathcal{V},\mathcal{E}>$ & geo-social network (GeoSN)\\
		$\mathcal{V}=\{U,L\}$ & node set in $\mathcal{G}$\\
		$U$ & user node set\\
		$L$ & POI node set\\
		$\mathcal{E}$ & $\mathcal{E}=\{\mathcal{E}_{u,u},\mathcal{E}_{u,l},\mathcal{E}_{l,l}\}$, edge set in the GeoSN $\mathcal{G}$\\
		$\mathcal{E}_{u,u}$ & set of social links\\ $\mathcal{E}_{u,l}$ & set of check-in links\\
		$\mathcal{E}_{l,l}$ & set of the association relationship between POIs\\
		$u_T$ & the target user \& activity  initiator\\
		$(u,x,\ell_x)$ & user activtiy\\
		$\mathcal{G}(U)$& $\mathcal{G}(U)=<U,\mathcal{E}_{u,u}>$, social network\\
		$k$ & core number\\
		$K$ & size of the recommended activity location list\\
		$h$ & size of the recommended user group\\
		$\alpha$ & attention weights (group evaluation index)\\
		$U^k$ & user set of $\mathcal{G}(U^k)$ satisfying the $k$-core constraint\\
		$\mathbf{U}^k$ & user group set $\mathbf{U}^k=\{U^k_j\}$\\
		$U^k_h$ & $h$-sized  user set $U^k$ in the connected $\mathcal{G}(U^k_h)$ \\
		$T$ & activity topic \& POI category\\
		$CUG$ & candidate user group set $CUG=\{U^k_{(i)}\}_{i=1}^n$\\
		$U^{\star}$ & recommended user group\\
		$L(U^{\star})$ & current location set of users in $U^{\star}$\\
		$T^{\star}$ & recommended activity topic\\
		$\ell_{T^{\star}}$ & recommended activity location corresponding to $T^{\star}$\\
		\hline\hline
	\end{tabular}
\end{table}

The differences between the two types of datasets are:
\begin{itemize}
    \item LBSNs do not contain explicit group information while the  EBSNs do;
    \item LBSNs contain more precious POI information, while the EBSNs record the location and event id without semantic information.
\end{itemize}

\section{Methodology}
\label{ASG}
The multi-typed geo-social group recommendation problem is addressed by solving three sub-problems: (1) how to obtain recommended user groups; (2) given the recommended user group, how to obtain the corresponding activity topics; (3) how to select the activity location based on the recommended activity topic.

\subsection{Obtaining Recommended User Groups and Corresponding Activity Topics}
\subsubsection{Algorithm Overview}
In this subsection, we describe how to address the first two sub-problems. The proposed scheme is termed as \textbf{social group-category query}.

In the social group-category query, we propose to explore candidate user group by adopting the $k$-core decomposition algorithm and breath-first-search (BFS). The procedure is composed of two steps: 
(1) obtaining the latent candidate user group set $\mathbf{U}^{k}$ by using core decomposition algorithm; 
(2) selecting the candidate user group set $\{ U_{(1)}^k, U_{(2)}^k,...\}, |U_{(i)}^k|=h$ from the set $\mathbf{U}^k$. 
User groups with $k$-core are not final recommendation results, and our goal is to retrieve a user group satisfying the following constraints: 
\begin{itemize}
    \item the retrieved user group contains the target user $u_T$;
    \item the users in the retrieved user group are friends of the target user;
    \item the size of the retrieved user group is $h$;
    \item the retrieved user group is a connected graph. 
\end{itemize}
We term the user groups that satisfy the constraints mentioned above, as \textbf{candidate user group $U^{k}_h$}.

The detailed algorithm of the social group-category query is illustrated in Algorithm \ref{alg1}. Specifically, sub-problem (2) is addressed by invoking Algorithm~\ref{alg2}.

\noindent \textbf{Algorithm~\ref{alg1}} first initializes the latent candidate user group $U^k$ with $k$-core, The connected graph $\mathcal{G}(U^k)$ spanned by $U^k$ is initialized to $H$. $U_c^k$ denotes the user set in the connected graph. We employ the core decomposition algorithm \cite{Batagelj} to detect the user group $U_c^k$ (line 3$\sim$9),  which satisfy the condition that the selected candidate user groups on $\mathcal{G}(U^k)$ are of size-$h$. According to the definition of $k$-core, the cardinality of $k$-core is set to be no less than $k+1$. Therefore, the algorithm enumerates user groups with cardinality in a gradually increasing order within the range of $[k+1, M]$. Algorithm~\ref{alg1} invokes Algorithm~\ref{alg2} for obtaining the recommended 2-tuple $\{U^{\star}, T^{\star}\}$, the input of Algorithm~\ref{alg2} is the user set $U_c^k$, $k$ and $h$. 

\noindent \textbf{Algorithm~\ref{alg2}} solves the sub-problem (2). It initially sets the final recommended user group $U_h^k$ to $\emptyset$, and the algorithm maintains two priority queues $UG$ and $DG$, as shown in line 2. $UG$ stores the unextended user groups (the user group that has not been selected yet). $DG$ manages the extended user groups, where store nodes selected based on BFS other than the randomly selected node in the initial stage. In line 3$\sim$11, users in $\mathcal{G}(U^k)$ are stored in $DN$ if they are neither included in any other user group nor stored in $UG$ and $DG$.

In lines 12$\sim$13, Algorithm~\ref{alg2} checks if $U_{(i)}^k$ is a connected $k$-core component as long as its size reaches $h$. 
In line 14$\sim$16, Algorithm~\ref{alg2} invokes the attention mechanism (\textbf{AttentionSelect} 
in the latter part of this section) to find the user group $U^{\star}$ with the maximal attention score. Finally, by calculating the predicted score with \emph{Bayesian Personalized Ranking} (BPR) pair-wise learning objective, the final recommended user group $U^{\star}$'s corresponding activity topics and $U^{\star}$ are returned as the results of Algorithm~\ref{alg1}.

\noindent \textbf{Complexity Analysis:} 
In this part, we detail the complexity analysis of Algorithm~\ref{alg1}. Firstly, according to \cite{Batagelj}, it is known that the time complexity of the core decomposition algorithm is $O(|U|+|\mathcal{E}_{u,u}|)$. Secondly, obtaining the candidate user group by Algorithm~\ref{alg2} has a time complexity of $O(\overline{d}^h \cdot h^2)$, where the parameter $\overline{d}$ represents the average degree of nodes in the given GeoSN $\mathcal{G}$. Lastly, packing an $h$-sized user group containing the target user $u_T$ in Algorithm~\ref{alg1} has a time complexity of $O(\overline{d}^h)$. 
In summary, the time complexity of Algorithm~\ref{alg1} is $O(|U|+|\mathcal{E}_{u,u}|+\overline{d}^h \cdot h^2)$.

\renewcommand{\algorithmicrequire}{\textbf{Input:}} 
\renewcommand{\algorithmicensure}{\textbf{Output:}} 
\begin{algorithm}[!t]
\caption{Social group-category query} \label{alg1}
\begin{algorithmic}[1]
\footnotesize
  \REQUIRE A geo-social network $\mathcal{G}$ and its subgraph: social network $\mathcal{G}(U)$, one targe user $u_T$ and his/her preference topic list $T_{u_T}=\{t_1,t_2,...\}$, a social group-category query $Q'=<u_T,h,k>$; $k$ and $h$ are  integers, $k$ is the core number of any vertex in one graph; $h$ is the group size.
  \ENSURE A group-category pattern $\mathcal{P}=(U^{\star},T^{\star})$.
 
  \STATE Initialize $U^k \leftarrow \emptyset$;
  \STATE Initialize $H \leftarrow$ All connected components of $\mathcal{G}(U^k)$
  \STATE $M\leftarrow \max_{U_c^k\in H}|U_c^k|$;
  \STATE \textbf{for} $h$ from $k+1$ to $M$ \textbf{do}
  \STATE \quad \textbf{for} each $U_c^k$ in $H$ \textbf{do}
  \STATE \quad \quad \textbf{if} $|U_c^k|\leq h$ \textbf{then}
  \STATE \quad \quad \quad $\{U_h^k, T_{U_h^k}\} \leftarrow$ \textbf{GetCandidateGroup}$(U_c^k, k, h)$;
  \STATE \quad \quad \quad \textbf{if} $U_h^k \neq \emptyset$ \textbf{then}
\STATE \quad \quad \quad \quad  \textbf{Return} $U^{\star} = U_h^k, T^{\star}=T_{U_h^k}$;
\STATE \textbf{Return} $\emptyset$
\end{algorithmic}
\end{algorithm}

\subsubsection{Recommended User Group $U^{\star}$ Selection}
The activity initiator (i.e., the target user $u_T$) is more likely to choose the users who respond to his/her call of activity. Our observations on the premises of successful group activity planning are:
\begin{itemize}
    \item \textbf{Premise~1}: participating users are interested in the activity topic;
    \item \textbf{Premise~2}: $u_T$ is an expert or experienced about the activity topic.
\end{itemize}
Based on the premises, we utilize the attention mechanism to evaluate the $u_T$'s decision-making power in the group. We also use the activity topic-aware influence of $u_T$ on each user in the candidate user group to represent the activity topic vote of each user with respect to the target user $u_T$. We select a user group as the recommended (optimal) user group, in which the target user $u_T$ is the main decision-maker about his/her familiar activity topics.

The architecture of the proposed social group-category query is illustrated in Figure~\ref{fig:AGC}. For the candidate user group set with $k$-core $CUG=\{U_{(j)}^k\}_{j=1}^n$ , each candidate user group $U_{(j)}^k$ is treated as one attention sub-network (ASN). The inputs of each ASN $U_{(j)}^k$ are the user-category vector $\mathbf{C }_j$ (yellow circles) and the set of user-latent vectors $\{\mathbf{u}_{(1)},\mathbf{u}_{(2)},...,\mathbf{u}_{(h-1)}\}$. It returns the attention weight $\alpha(j,u_{(i)},u_T)$ for any $u_{(i)} \neq u_T, u_{(i)} \in U_{(j)}^k$ (green rectangles). 

$U_{(j)}^k$, the output of each ASN, is calculated by cumulative score sum $\alpha_{u_T}(U_{(j)}^k)=\sum_{u_{(i)} \neq u_T}^{h}\alpha(j,u_{(i)},u_T)$ (blue rectangle). In this paper, we utilize the hard attention mechanism to select the user group with the maximum score (white rectangle); thus, the recommended user group $U^{\star}$ can be derived as:
\begin{equation}
\small
U^{\star}=\arg \max_{i=1}^{n}\alpha_{u_T}(U_{(i)}^k).
\end{equation}

In each candidate user group $U_{(j)}^k$, the preference votes of $u_i$ on the other users is represented by attention weight in each ASN model. At the same time, the interactions between user $u_i \in \forall U_{(j)}^k$ and $u_T$ is modelled by the ASN.
 
The attention score $a(j, u_{(i)}, u_T), u_{(i)} \in U_{(j)}^k$ of each candidate user group in $CUG$ is can be derived by:
\begin{equation}
a(j,u_{(i)}, u_T)= \mathbf{w}^T\sigma(\mathbf{C}_{u_{(i)}},\mathbf{u}_{u_T,u_T\neq u_{(i)}})+\mathbf{c}, \quad i=[1,h]
\end{equation}
where $\mathbf{C}_{U_{(j)}^k}$ is the user-category vector and $\{\mathbf{u}_{(1)},\mathbf{u}_{(2)},...,\mathbf{u}_{(h)}\} \smallsetminus \{\mathbf{u}_{T}\}$ is a set of user-latent vectors.

In this paper, influence function is used to represent the influence gain from the target user $u_T$ on the user $u_{(i)}$. We define $\sigma(\cdot,\cdot)$ as the following equation.
\begin{equation}
\small
\begin{split}
&\sigma(u_T,u_{(i)})=\\ &\begin{cases} \frac{\sum_{t\in T_{u_T} \cap T_{u_{(i)}}}p(t,u_T)\cdot p(t,u_{(i)})}{\sqrt{\sum_{t\in T_{u_T}}p(t,u_T)^2 }\sqrt{\sum_{t\in T_{u_{(i)}}} p(t,u_{(i)})^2}}, & if\; p(t,u_T)\geq p(t,u_{(i)});\\
0, & otherwise.
\end{cases}
\end{split}
\end{equation}
The final attention weights are derived by normalizing $a(j, u_{(i)}, u_T)$ using Softmax function.
\begin{equation}
\begin{split}
&\alpha(j,u_{(i)},u_T)=\frac{exp(a(j,u_{(i)},u_T))}{\sum_{j=1,u_{(i)}\neq u_T}^h exp(a(j,u_{(i)},u_T))};\\
&\alpha_{u_T}(U_{(j)}^k)=\sum_{u_{(i)} \neq u_T}^{h}\alpha(j,u_{(i)},u_T), \forall u_{(i)} \in U_{(j)}^k.
\end{split}
\end{equation}

\subsubsection{Recommended Activity ($T^{\star}$) Selection using AttentionSelect}

\begin{figure*}[!t]
\centering
\includegraphics[width=0.8\linewidth]{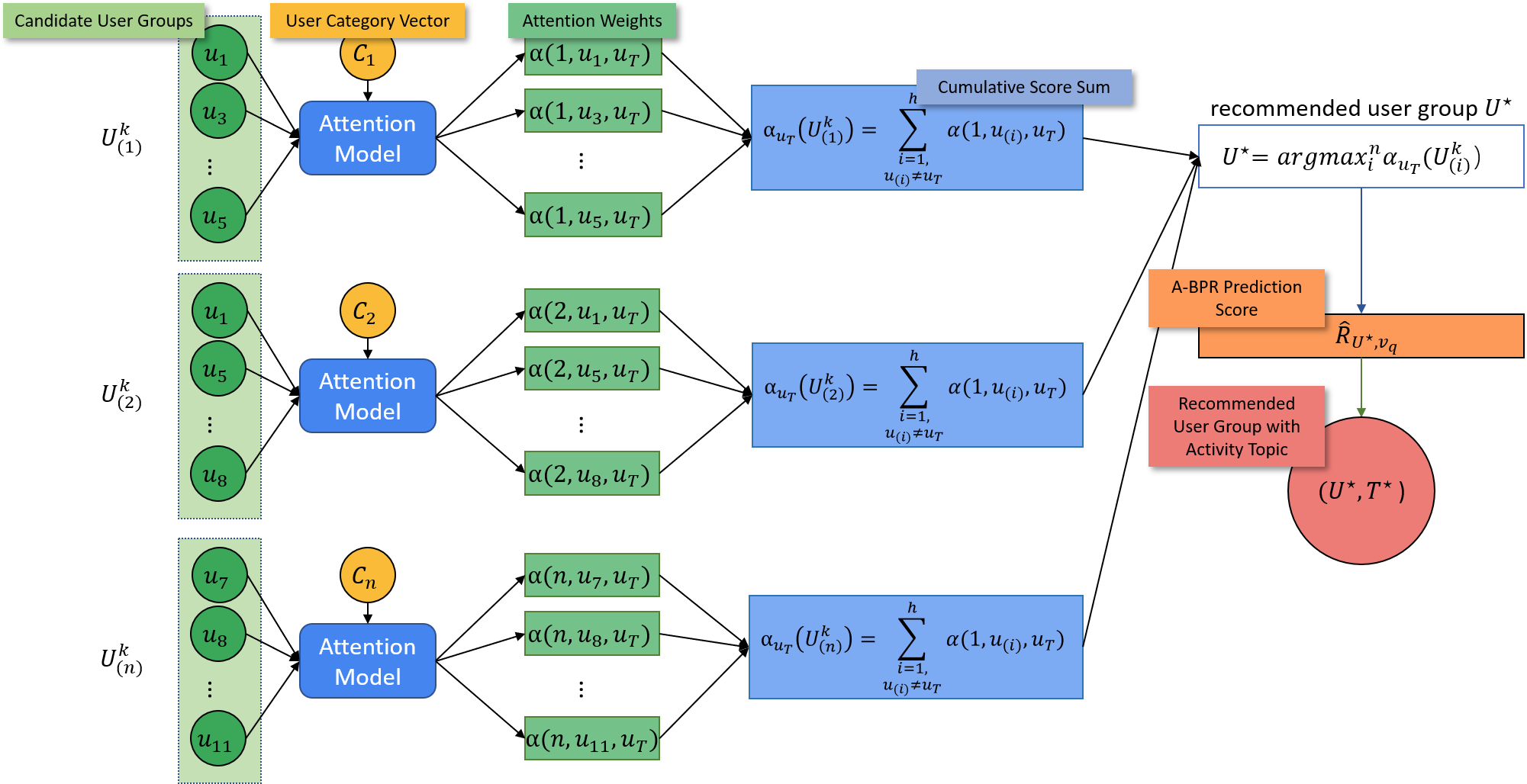}
\caption{An overview of social group-category query.} 
\label{fig:AGC}
\end{figure*}

As illustrated in Figure~\ref{fig:AGC}, after obtaining $U^{\star}$, the remaining task of \textbf{AttentionSelect} is to find the recommended activity topic $T^{\star}$ for $U^{\star}$. The target user should have the strongest decision-making power in $U^{\star}$ under topic $T^{\star}$. Specifically, we introduce attention mechanisms to BPR (termed as A-BPR) and use A-BPR to select recommended activity topic (orange rectangle). The location categories visited by users in $U^{\star}$ and the location categories that are not visited (partial users have visited) are combined randomly into a set of 2-tuples. The categories in the 2-tuple are scored by A-BRP to find a top-1 activity topic $T^{\star}$ (red circle).

The pair-wise ranking between visited and unvisited location categories is conducted by A-BPR in the social group-category query. The scoring function and optimization objective function for A-BPR ranking are described in the following paragraphs.

\noindent \textbf{Predicted Score:} 
Given the selected recommended user group $U^{\star}$, the predicted score $\widehat{R}_{U^{\star},v_q}$ for $U^{\star}$ and visited location category $v_q$ can be derived as follows:
\begin{equation}
\small
\widehat{R}_{U^{\star},v_q}=\alpha_{u_T}(U^{\star})\cdot\mathbf{u_T}^T \mathbf{v}_q = \sum_{u_{(i)} \neq u_T}^{h}\alpha(U^{\star},u_{(i)},u_T)\cdot\mathbf{u_T}^T\mathbf{v}_q
\end{equation}
where $\mathbf{v}_q$ is the item latent vector for activity topic $v_q$.

\noindent \textbf{Objective Function:} 
The objective function of A-BPR pair-wise ranking is formulated as follows:
\begin{equation}
\small
\arg \min_{\Theta} \sum_{(j,v_q,v_s)\in \mathbb{D}_{\star}} -\ln\varphi(\widehat{R}_{U^{\star},v_q}(\Theta)-\widehat{R}_{U^{\star},v_s}(\Theta))-\lambda\Vert\Theta\Vert^2
\end{equation}

Substituting Equation (5) into Equation (6), the objective function can be expressed as:
\begin{equation}
\label{f3}
\small
\begin{split}
&\arg \min_{\Theta} \sum_{(U^{\star},v_q,v_s)\in \mathbb{D}_{\star}}-\ln \varphi\{(\sum_{u_{(i)} \neq u_T}^{h}\alpha(U^{\star},u_{(i)},u_T)\cdot\mathbf{u_T}^T\mathbf{v}_q\\&-(\sum_{u_{(i)} \neq u_T}^{h}\alpha(U^{\star},u_{(i)},u_T)\cdot\mathbf{u_T}^T\mathbf{v}_s\}+\lambda(\Vert\Theta\Vert)^2 
\end{split}
\end{equation}
, where the set $\mathbb{D}_{\star}$ contains all pairs of the visited and un-visited location categories for the recommended user group $U^{\star}$; $\Theta$ denotes the model parameters; $\alpha(U^{\star},u_{(i)},u_T)$ weights the vote of user $u_{(i)}$ for user $u_T$ in the user group $U^{\star}$. Adaptive Moment Estimate (Adam) is used to optimize the objective function. Specifically, $\varphi$ is the logistic sigmoid function; $\lambda$ is the regularization parameter.

\subsection{Obtaining Recommended Activity Location}
\label{AGSG}

\renewcommand{\algorithmicrequire}{\textbf{Input:}} 
\renewcommand{\algorithmicensure}{\textbf{Output:}} 
\begin{algorithm}[!t]
\footnotesize
\caption{GetCandidateGroup} \label{alg2}
\begin{algorithmic}[1]
  \REQUIRE All connected components $H$ of $\mathcal{G}_U(U^k)$, $h$ is the size constraint of the candidate user group.
  \ENSURE $\{U_h^k,T_{U_h^k}\}$.
  \STATE Initialize $U_h^k \leftarrow \emptyset$;
  \STATE Initialize priority queue $UG \leftarrow u_T, DG \leftarrow \emptyset, DN\leftarrow \emptyset$;
\STATE \textbf{while} $UG \neq \emptyset$ \textbf{do}
\STATE \quad $U_{(i)}^k \leftarrow UG.dequeue()$;
\STATE \quad  $|U_{(i)}^k |<h$ \textbf{then}
\STATE \quad \quad Select $u \in U_{(i)}^k $ with highest degree and $u \notin DN$;
\STATE \quad \quad $DN \leftarrow DN \cup u$;
\STATE \quad \quad \textbf{for} each set $Vp \subseteq
N(u)/U_{(i)}^k , k \leq |Vp|+deg_{\mathcal{G}_U(U_{(i)}^k )}(u)$ \textbf{and} $|Vp|+|U_{(i)}^k |\leq h$ \textbf{do}
\STATE \quad \quad \quad induce a subgraph $G_i$ of $U_{(i)}^k  \cup Vp$ in all connected components of $\mathcal{G}_U(U^k)$;
\STATE \quad \quad \quad \textbf{if} $G_i \nsubseteq UG \cup DG$ \textbf{then}
\STATE \quad \quad \quad \quad $UG \leftarrow UG \cup G_i$;\\

\STATE \quad \textbf{if}  $|U_{(i)}^k |=h$ and $U_{(i)}^k  \nsubseteq DG$ \textbf{then}
\STATE \quad \quad $DG \leftarrow DG \cup U_{(i)}^k $;
\STATE \quad \quad \textbf{if} $\{u\in U_{(i)}^k : k >|deg_{\mathcal{G}_U(U_{(i)}^k )}(u)|\}\neq \emptyset$ \textbf{then}
\STATE \quad \quad \quad $\{U_{(i)}^k ,T_{U_h^k}\} \leftarrow$ \textbf{AttentionSelect}$(U_{(i)}^k , \alpha_{u_T}(U_{(i)}^k ))$;
\STATE \textbf{Return} $\{U_h^k,T_{U_h^k}\}$
\end{algorithmic}
\end{algorithm}

Given the obtained recommended user group $U^{\star}$ and corresponding activity topic $T^{\star}$, in this section, we address the problem of finding an optimal activity location by exploring the spatial information. 

As described in the previous section, Algorithm~\ref{alg1} returns the activity topic (location category). However, there are many locations (POIs) in the geography space that match the corresponding activity topic. An intuitive approach is to perform the activity topic filtering operation on activity topics returned from Algorithm~\ref{alg1}. Therefore, we select the locations under the recommended activity topic as the input of the activity location query.

In this paper, aggregate nearest neighbour (ANN) query is introduced to perform the activity location query. ANN query reports the activity location that minimized the maximum distance that each user in $U^{\star}$ has to travel, which causes all users in the earliest time $U^{\star}$to reach the activity location \footnote{we assume that they move with the same speed}. The aggregate distance between the activity location $l$ and the user locations $L(U^{\star})$ can be denoted by the following equation:
\begin{equation}
adist(l_i,L(U^{\star}))=\max_{i=1}^{m}|l_il.u_j|, \forall u_j \in U^{\star}
\end{equation}
where $|l_il.u_j|$ is Euclidean distance between two locations $l_i$ and $l.u_j$, $l.u_j$ denotes $u_j$'s current location, $l_i$'s category belongs to the recommended activity topic $T^{\star}$.

\noindent \textbf{Algorithm Design:}
A single point approach (SPA)\cite{Papadias} is introduced to process the aggregate nearest neighbour (ANN) query. The SPA  calculates the aggregated centroid $l_{cen}$ of all users' current location set $L(U^{\star})$, and the requirement of the aggregated centroid $l_{cen}$ is to ensure that the value of  $adist(l_{cen},L(U^{\star}))$ is minimal.

It is worth noting that the execution of the centroid calculation is in the initialization phase of the SPA, and that any I/O operations are not generated during the centroid calculation process because of $L(U^{\star})$ being memory-resident.

\noindent \textbf{The Centroid-based Pruning:} 
To accelerate the ANN query, we propose a centroid-based pruning method to prune the search space of ANN query. Finding the aggregated centroid can be reduced to a \emph{minimum enclosing circle problem}. There are currently many effective algorithms to solve this problem. Since the randomized incremental algorithm has a lower time complexity (linear) relative to the number of query points, it is cited herein to solve the problem. However, it is worth noting that there is currently no better way to accurately calculate the centroid within the area consisting of the weighted locations. Therefore, we can use the same centroid for both weighted and unweighted cases.

After aggregate centroid is calculated, we provide lemmas for pruning the search space.
\newtheorem{lemma}{Lemma}
\begin{lemma}
\label{lema1}
Let $L(U^{\star})=\{l.u_{(1)},l.u_{(2)},...,l.u_{(h)}\}$ be a set of query locations,  and $l.u$ is an arbitrary user's current location in the geography space. 
Any location $\ell$ in the geographic area where $L(U^{\star})$ is located that belongs to the active topic $T$ applies to the following inequality: $adist(\ell,L(U^{\star})) \geq f(|\ell l.u|-|l.u_{(1)}l.u|,...,|\ell l.u|-|l.u_{(h)}l.u|)$, where $|\ell l.u|$ represents the Euclidean distance between $\ell$ and $l.u$.
\end{lemma}

\noindent \emph{proof:}
According to the triangle inequality, for each query location $ l.u _ {(i)} $, we get: $|\ell l.u_{(i)}|+|l.u_{(i)}l.u| \geq |\ell l.u| \Longrightarrow |\ell l.{(i)}| \geq |\ell l.u|-|l.u_{(i)}l.u|$. Due to the monotonicity of $f$:
\begin{equation}
\small
\begin{split}
adist(\ell,&L(U^{\star}))=f(|\ell l.u_{(1)}|,...,|\ell l.u_{(h)}|)\\& \geq f(|\ell l.u|-|l.u_{(1)} l.u|,...,|\ell l.u|-|l.u_{(h)} l.u|).\quad \square
\end{split}
\end{equation} 

Based on Lemma~\ref{lema1}, we introduce one heuristic for pruning intermediate nodes. The intuition of the heuristic is to discard locations that cannot contain qualified ANNs.

\newtheorem{heuristic}{Heuristic}
\begin{heuristic}
\rm Let $l_{cen}$ is the centroid of $ L(U^{\star})$, so far $ best\_dist$ is the best ANN distance. All collections  $L_{T^{\star}}$ of locations belonging to the recommended activity topic $T^{\star}$ can be pruned if:
\begin{equation}
\begin{split}
\max_{i=1}^h&(mindist(L_{T^{\star}},l_{cen})-|l.u_{(i)}l_{cen}|)\geq best\_dist\Leftrightarrow \\& mindist\left(L_{T^{\star}},l_{cen}\right) \geq best\_dist + \min_{i=1}^h|l.u_{(i)}l_{cen}|.
\end{split}
\end{equation}
, Where $mindist(L_{T^{\ star}}, l_{cen})$ represents the minimum distance between the MBR of $L_{T^{\star}}$ and $l_{cen}$, and $best\_dist$ represents the aggregate distance.
\end{heuristic}

\renewcommand{\algorithmicrequire}{\textbf{Input:}} 
\renewcommand{\algorithmicensure}{\textbf{Output:}} 
\begin{algorithm}[!t]
\caption{SPA-DF} \label{alg3}
\begin{algorithmic}[1]
\footnotesize
  \REQUIRE The users' current location set $L(U^{\star})$, and its aggregate centroid $l_{cen}$, $L_{T^{\star}}$ is the set of locations with respect to the activity topic $T$ in the geography space.
  \ENSURE one location $\ell \in L_{T^{\star}}$ is the nearest distance to each user $u_i$'s current location in $U^{\star}$.
  \STATE \textbf{if} $l_j$ is an intermediate location 
  \STATE \quad sort location $l_j$ in $L'$ according to $mindist(l_j,l_{cen})$ in $list$
  \STATE \quad \textbf{repeat}
  \STATE \quad \quad get next location $l_j$ from $list$
   \STATE \quad \quad \textbf{if} $\max_{i=1}^h(mindist(L_{T^{\star}},l_{cen})-|l.u_{(i)}l_{cen}|)< best\_dist$
  \STATE \quad \quad \quad \textbf{SPA}$(l_j,L(U^{\star}))$
  \STATE \quad \textbf{until} $\max_{i=1}^h(mindist(L_{T^{\star}},l_{cen})-|l.u_{(i)}l_{cen}|)\geq best\_dist$ or end of $list$
  \STATE \textbf{else if} the location $l_j$ is a leaf node
  \STATE \quad \textbf{for} each location $\ell_j$ in $l_j$
  \STATE \quad \quad \textbf{if} $adist(\ell_j,L(U^{\star})) <best\_dist$
    \STATE \quad \quad \quad $best\_NN = \ell_j; best\_dist = adist(\ell_j,L(U^{\star}))$.
\end{algorithmic}
\end{algorithm}

Based on the analysis above, depth-first search is leveraged to complete the SPA. Algorithm~\ref{alg3} illustrates the pseudo-code of SPA-DF. For $L_{T^{\star}}$, the location in the list is sorted according to $minist$ from the centroid $l_{cen}$ and accessed in order (recursively).
When the first location $ l_j $ with $\max_{i=1}^h(mindist(L_{T^{\star}},l_{cen})-|l.u_{(i)}l_{cen}|) \geq best\_dist$ is found, the algorithm removes the subsequent ones in the list.

It's worth noting that because $|l.u_{(i)} l_{cen}|$ is a constant and $max$ is monotonous, so the order according to $mindist$ is the same as the order according to $\max_{i=1}^h(mindist(L_{T^{\star}},l_{cen})-|l.u_{(i)}l_{cen}|)$. The SPA is similar to the algorithm for retrieving the normal NN of $l_{cen}$; however, the difference between them is the aggregation-based condition.

To conclude this section, we prove the following lemma as a proof of algorithm correctness based on \cite{Papadias}:
\begin{lemma}
\label{lema2}
The SPA is able to correctly report the location in $L$  with minimum aggregate distance from the $L(U^{\ star})$ .
\end{lemma}

\noindent \emph{proof:}
\; It is sufficient to prove that Heuristic 1 mentioned in this section is safe. It can also be equivalently stated that the locations of the SPA method pruning do not contain  better NN than the one that has been found. For any location $l$ in $L$, it applies to $|ll_{cen}|\geq mindist(L,l) \Rightarrow |ll_{cen}|-|l.u_{(1)}| \geq mindist(L,l_{cen})-|l.u_{(i)}l_{cen}|$. Thus, due to the monotonicity of the function，if and only if $\max_{i=1}^h(mindist(V,$ $l_{cen})-|l.u_{(i)}l_{cen}|)\geq best\_dist$. 

According to Lemma \ref{lema1}, we can get $adist(l,L(U^{\star})) \geq best\_dist$ for any location $l$ in $L$, so it can be proved that $L$ can be safely pruned. 
The whole aforementioned inequalities (including Lemma \ref{lema1}) apply to positive weights; therefore, the proof of correctness includes weighted functions.
$\square$

\section{Experiments}
\subsection{Dataset Description.}
In this paper, we select two kinds of geo-social networks dataset: location-based social network (LBSN) datasets (e.g., Foursquare\cite{Yin}, Gowalla\footnote{http://snap.stanford.edu/data/}); event-based social network (EBSN) datasets (e.g.,Yelp \cite{Yin}, Plancast used in \cite{LiuX}) for the observation and evaluation. Table~\ref{tab1} shows the basic statistics of the data set in the experimental section.

\begin{table}[htbp]
\footnotesize
\caption{Statics of the Datasets}
\label{tab1}
	\centering
	\begin{tabular}{c|cc|cccc}
		\hline\hline
		\textbf{Dataset}&\textbf{Foursquare}&\textbf{Gowalla}&\textbf{Yelp}&\textbf{Plancast} \\ 
		\hline\hline
		\# Users&3,305&5,623&$34,504$&$ 72,834$\\
		\# Group Events&105,470&406,333&$22,212$&$207,342$\\
		\# Items&105,470&406,333 &$8,016$&$113,361$\\
		\hline\hline 
		Avg. Group Size &25.191&28.259&$11.504$&$12.027$\\
		Avg. \#Items for   &1&1&$4.60$&1\\ a Group \\
		Avg. \#Record for &115.441&231.193&$3.70$&$9.343$\\ a User \\
		Avg. \#Friends for  &7.812& 9.668&$3.088$&$4.295$\\ a User\\
		Avg. \# Records for &2.185e+04&3.199&$ 2.315$&$2.164$\\an Item \\
		\hline\hline
	\end{tabular}
\end{table}

The differences between the two types of datasets are:
\begin{itemize}
    \item LBSNs do not contain explicit group information while the  EBSNs do;
    \item LBSNs contain more precious POI information, while the EBSNs record the location and event id without semantic information.
\end{itemize}
We extract implicit group check-ins using the following identification criteria: \emph{if the target user and his / her friends check-in at the same venue (POI) on the same check-in time, we regard them as an activity group}. More specifically, check-in records will be treated as a group check-in record as long as $|t_t - t_f| < 1800 (seconds)$, where $t_t$ and $t_f$ are target user check-in time and friend check-in time respectively. 

\subsection{Baseline Methods}
Most previous related researches focus on recommending single-typed recommendation content. To our best knowledge, no comparable work can be applied directly to solve the multi-typed recommendation problem proposed in this paper. Therefore, we build the baselines by combining several existing relevant methods as well as changing/removing part of AGSGR method. The experiments on latter methods also serve as an ablation study.

For the user group in the recommendation result, we set three baseline methods: \emph{\textbf{baseline1:} AGSGR\_GQ$^{\star}$}, proposed AGSGR without enforcing the social constraints between the target user and the other users in the group; \emph{\textbf{baseline2:} RS}, selecting a user group randomly; \emph{\textbf{baseline3:} GEM-E}, GEM method \cite{Hongzhi} enhanced by taking $h$ partners of the same event as the recommended user group\footnote{The output of GEM is a 2-tuple \emph{(partner, event)}}.

When the recommended user group information is obtained, existing state-of-the-art group recommendation approaches are introduced to make recommendations on activity locations using the retrieved user group information. These group recommendation methods are PIT \cite{Liu}, COM \cite{Yuan} and AGR \cite{Tran}. PIT is a topic-based  probabilistic graphical model. 
COM assumes that the personal influence is dependent on the topic, and the topic-aware/personal preferences of the group affect the final decision. AGR combines attention sub-networks and BRP to create an attentive model that learns the dynamic personal impact weights of each user to make group recommendations. Being different from AGSGR method, which assumes that the initiator (the target user $u_T$) is the main decision-maker in an activity, AGR ignores the analysis and selection of activity topics.

In summary, since the existing group recommendation method assumes that the group is known, we combine the existing group recommendation methods with geo-social group query methods as baselines in our experiments. 
In our experiments, we mainly refer to the parameter settings in the paper \cite{Tran} regarding the parameter settings of the baseline methods.

\begin{figure*}[!t]
\centering
\subfigure[Precision of the group query in Foursquare]{
\begin{minipage}[b]{1\textwidth}
\includegraphics[width=1\textwidth]{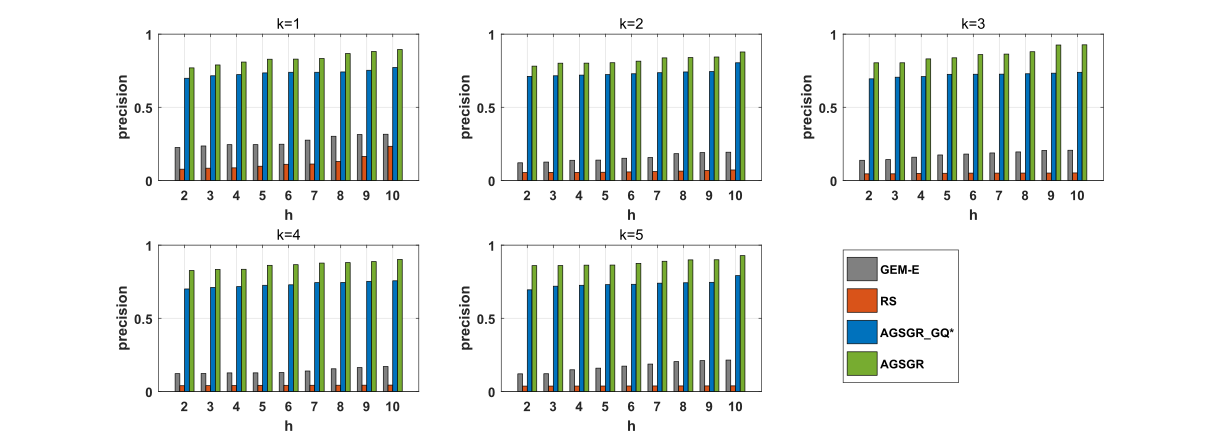}
\end{minipage}}
\subfigure[Precision of the group query in Gowalla]{
\begin{minipage}[b]{1\textwidth}
\includegraphics[width=1\textwidth]{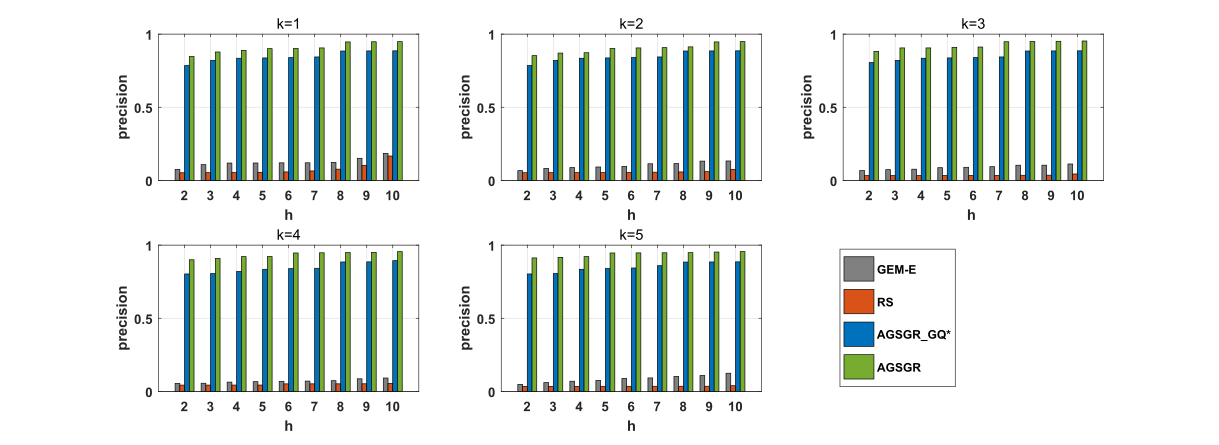}
\end{minipage}}
\caption{The performance of group query in Foursquare and Gowalla datasets (AGSGR \& AGSGR\_GQ$^{\star}$, RS and GEM-E with different value of $K$ and $h$.} 
\label{fig:1}
\end{figure*}

\begin{figure*}[!t]
\centering
\subfigure[Precision of the group query in Plancast]{
\begin{minipage}[b]{1\textwidth}
\includegraphics[width=1\textwidth]{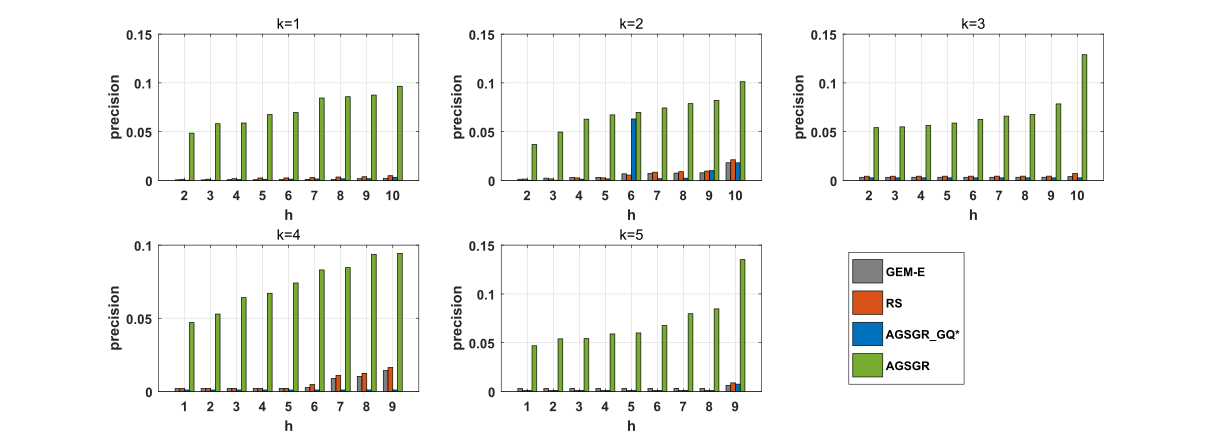}
\end{minipage}}
\subfigure[Precision of the group query in Yelp]{
\begin{minipage}[b]{1\textwidth}
\includegraphics[width=1\textwidth]{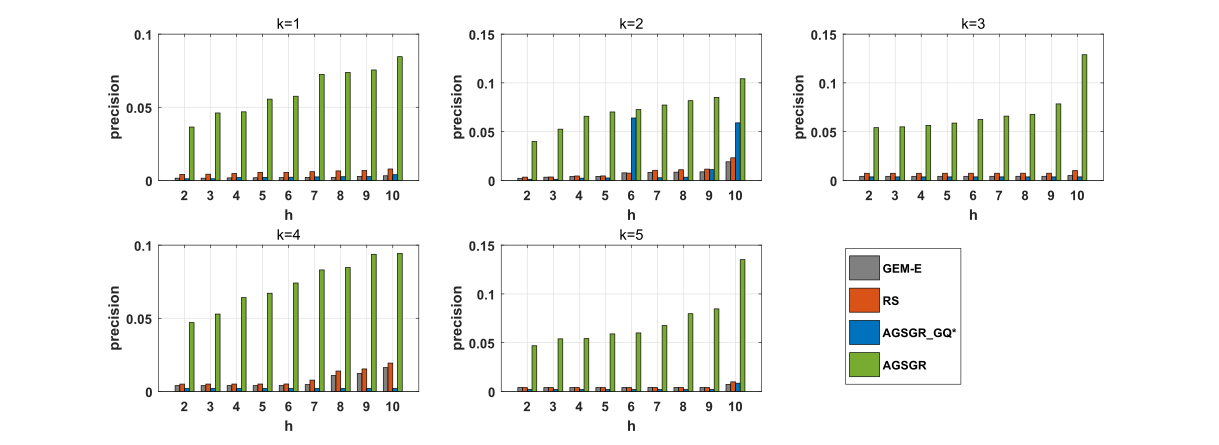}
\end{minipage}}
\caption{The performance of group query in Plancast and Yelp dataset (AGSGR \& AGSGR\_GQ$^{\star}$, RS and GEM-E with different value of $K$ and $h$.} 
\label{fig:2}
\end{figure*}

\subsection{Performance Metrics.}
According to the problem definition  proposed in this paper, the recommendation results of the multi-typed group recommendation are:
\begin{itemize}
    \item top-$M$ user group list, $U_i^{\star} \in \{U_{(1)}^{\star},U_{(2)}^{\star},...,U_{(M)}^{\star}\}$
    \item $M$ activity locations $L_{Rec}(U_i^{\star})$ corresponding to user group $U_i^{\star}$
\end{itemize}
Due to the fact that the recommendation results are heterogeneous, we evaluate the performance of AGSGR in terms of activity group prediction accuracy and group activity location prediction accuracy.

\subsubsection{Evaluating the performance of activity group prediction}
For activity group prediction, we use precision to measure the accuracy of the recommendations. The precision rate represents the percentage of user group returned by the $h$ recommended user group who belong to the real activity group in the real dataset with the activity groups. Formally, it is denoted by: 
\[
Pre@h=\frac{1}{N}\sum_{i=1}^{N}\frac{|U^{\star}_{(i)}\cap U_{tru}|}{|U^{\star}|}, \;|U^{\star}_{(i)}|=h
\]
, where $N$ is is the number of target users randomly sampled from each dataset, $U^{\star}$ is the recommended user group;  $U_{tru}$ is the real group information contained in the LBSN or EBSN datasets. $h$ is the size of the recommended user group as well as the constraint of the group size. For experiments in this paper, $N$ is set to $100$, the value of $k$ and $h$ are set to be the range of $[1,5]$, $[2,10]$, respectively.

\subsubsection{Evaluating the performance of activity location prediction}
For activity location prediction, we focus on the activity location prediction based on the activity groups obtained above.
This paper utilizes two metrics to demonstrate the performance of proposed method and baselines: $Precision@K$ \cite{Yuan}\cite{Zhao} and normalized discounted cumulative gain $nDCG@K$ \cite{Manning}, where $K$ is the number of recommended activity locations (POIs).

The $Precision@K$ represents the percentage of the activity locations returned by the top $K$ recommended locations adopted by a group. It can be calculated using the following equation: 
\[
Pre@K = \frac{1}{M} \sum_{j=1}^M\frac{|L_{Rec}(U^{\star}_{(j)})\cap L_{True}(U^{\star}_{(j)})|}{|L_{Rec}(U^{\star}_{(j)})|}
\]
where $M$ is the number of recommended user groups obtained in the first step of the geo-social group recommendation, $L_{Rec}(U^{\star})$ denotes the top $N$ recommended activity locations for the recommended group $U_j$, and $L_{True}(U^{\star})$ denotes the ground truth activity locations for the recommended group $U^{\star}_{(j)}$.

Since the recommendation results are ranked, the normalized discounted cumulative gain $nDCG@K$ \cite{Manning}\cite{Yuan}\cite{Zheng} is introduced to measure how well a method can rank the true item higher in the recommendation list. It is denoted by the following equation: 
\[
nDCG = \frac{DCG}{MaxDCG}
\]
where 
\[
DCG = rel_1+\sum_{i=2}^K \frac{rel_i}{\log_2(i)}
\]
and $rel_i$ indicates whether the $i$-th location (POI) in each user group's activity location recommendation list is accepted by the group. If accepted, $rel_i=1$; otherwise,  $rel_i=0$. The size of the activity location recommendation list is $N$. $MaxDCG$ represents the maximum possible discounted cumulative gain $nDCG@K$ with optimal top $K$ recommendations, which normalizes the $DCG$. 

For each testing group, a $nDCG$ value is evaluated.  The $nDCG@K$ metric is then computed by averaging all $nDCG$ values obtained.

\subsection{Performance comparison}
\subsubsection{User Group Recommendation}
In this subsection, we mainly concentrate on performance comparison in both LBSNs and EBSNs. The performance of the proposed methods and the baselines on LBSN is demonstrated in Figure~\ref{fig:1}(a) and Figure~\ref{fig:1}(b). As illustrated in the figures, it is obvious to find that AGSGR  dominates the baselines (AGSGR\_GQ$^{\star}$, RS and GEM-E). 

The comparison between the proposed method and AGSGR\_GQ$^{\star}$ confirms the necessity of introducing the social constraint that the recommended users have a social connection with the target user. Since the major difference between the two methods is that the AGSGR\_GQ$^{\star}$ does not enforce the constraint mentioned above, AGSGR\_GQ$^{\star}$ outputs user groups that including both friends and non-friends of the target user. However, most people tend to join activities with people that have a close relationship (which is often represented as friendship on an LBSN) with them. As a consequence, AGSGR\_GQ$^{\star}$ exhibits lower recommendation accuracy. It is also worth mentioning that removing social constraints makes AGSGR\_GQ$^{\star}$ more suitable for the cold-start problem in the recommendation.

RS randomly select $h$ users as the candidate user group. The probability that the selected users include the target user's friend is even lower. It does not guarantee to comply with the social constraint mentioned above, and it does not take measure to optimize the target user's influence on the candidate user group either. Therefore, it demonstrates significant performance disadvantages in the comparison.

Though GEM-E is not designed to follow the social constraint when it recommends the candidate user group, it recommends the candidate user group based on the users' preference for the activity topic. For a target user in LBSN, the similarity between he/she and the other users determines how likely they will be included in the candidate user group. 

GEM-E select the top-$h$ users based on the similarity; however, such heuristic ignores the decision-making process of planning activity. Yin et al.\cite{{Hongzhi}} argued that the target user's impact on the decision-making process is critical because people intend to have a different preference regarding personal activity and group activity. People tend to believe in one who has a rich experience in related activities. As a result, on the one hand, some of the user recommended by GEM-E may not accept the activity initiated by the target user simply because they do not believe in his decision, which lowered its performance when compared with AGSGR; on the other hand, the GEM-E outperforms RS, due to the fact that recommending based on similarity in topic preference may raise the probability that the recommended users comply with the proposed social constraints.
For the performance comparison of the proposed method and baseline methods on EBSN in term of user group recommendation, as illustrated in Figure~\ref{fig:2}(a) and Figure~\ref{fig:2}(b), AGSGR continues to show superiority over the baseline methods (AGSGR\_GQ$^{\star}$, RS and GEM-E).

A feature of EBSN records is that the target users usually have a relatively lower chance to have friends with them when joining a single EBSN event, so all the methods suffer from recommendation accuracy degradation as shown in the figures. For the proposed methods, since the recommended user maintains a high probability of hitting the users in the ground truth, it still dominates the comparison.

Apart from that, it can be observed that RS outperforms AGSGR\_GQ$^{\star}$ in the comparison, which is contradictory to the results shown in the previous subsection. Since RS selects user randomly, it recommends more non-friend users for the target user; thus, random selection exhibits better results (though neither of them performs well). It should be noted, for AGSGR\_GQ$^{\star}$, the performance reaches a peak when $k=2, h=6$. After inspection of the recommendation results, we find that the recommended users have similarity with users recommended by AGSGR, because the user group recommended in AGSGR\_GQ$^{\star}$ method satisfying 2-core constraint and other social constraints almost covers the recommended user group in AGSGR satisfying the social constraints. 
 
For GEM-E, it selects users from users' friends; however, it ignores the decision-making power of the target user. Consequently, it performs even worse than RS. 
\begin{figure*}[!t]
\centering
\includegraphics[width=0.88\textwidth]{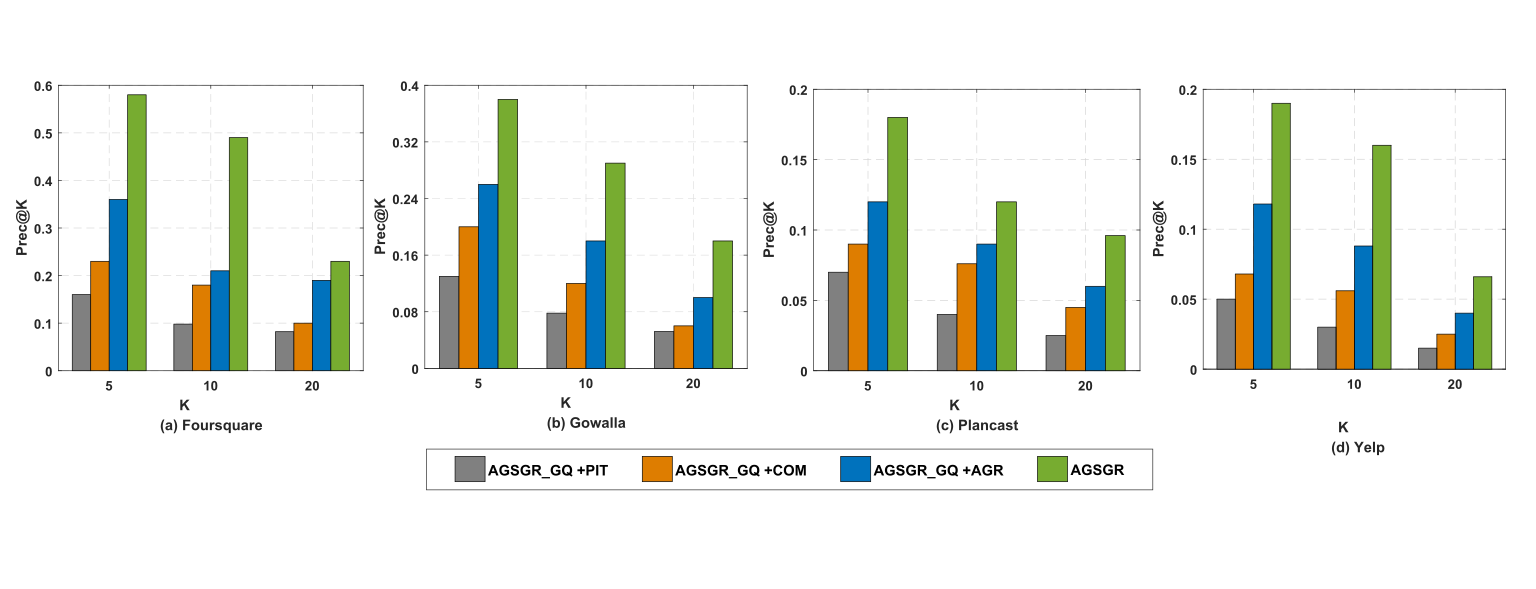}
\vspace{-12mm}
\caption{Prec@K evaluated based on obtained user group.}
\label{fig:3}
\end{figure*}

\begin{figure*}[!t]
\centering
\includegraphics[width=0.88\textwidth]{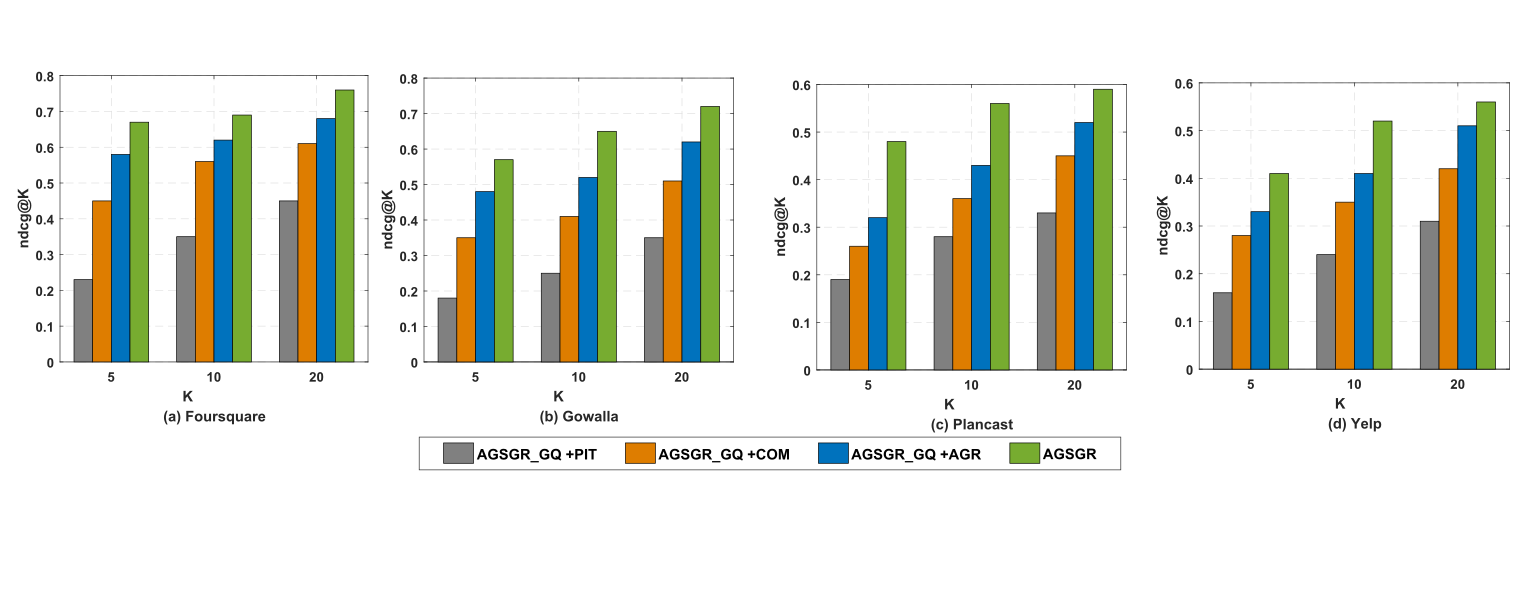}
\vspace{-12mm}
\caption{ndcg@K evaluated based on obtained user group.}
\label{fig:4}
\end{figure*}
\subsubsection{Activity Location Recommendation}
In this subsection, the results on activity location recommendation is presented. As shown in Figure~\ref{fig:3} and Figure~\ref{fig:4}, AGSGR again exhibit advantages in performance. It should be noted that the combination using RS is omitted since RS's performance on recommending the user group is too low to output any useful activity location recommendation. Therefore, we omitted the baselines using RS as the user group recommendation method.

AGR introduces an attention mechanism to discover the decision-maker in the existing user group and recommends the user group with the activity based on the preference and activity history of the decision-maker. Since the motivation of AGR is to find the decision-maker in the existing user group, which is different from the background of this paper, the decision-maker discovered by AGR is not necessarily the target user described in this paper. Therefore, the location recommendation results by AGR does not achieve the best performance. 

Both PIT and COM try to balance the preferences of all users in the user group, and the decision-making power of the target user is ignored either. They all assume that, for a certain user, the probabilities he/she would join an activity with different people are identical, because they presuppose that the probability of attending an activity is mainly determined by the target user, while different partners do not change the probability of participating in the event.
For PIT method, it assumes that the decision should be made by the user who has the greatest influence based on the social network graph. The most critical issue of this assumption is that the user with the greatest influence in the user group does not guarantee he/she is experienced in all the activity topics and can make the others trust him/her. 

The COM method shows advantages over the PIT method. Though the COM method assumes that the users' activity preference is related to the whether the user is willing to participate in group activities, it is common for the minority to follow the majority during the decision-making process. As a result, it has a relatively worse performance compared with the proposed method.

\section{Related Work}
Our research work is closely related to group recommendation, geo-social query, and event-partner recommendation. 

\subsection{Group Recommendation}
Group recommendation recommends a user group with items that attract them. 
Memory-based approach and model-based approach are two different approaches to achieve group recommendation\cite{Yahia}. 

Memory-based methods are realized by either preference aggregation \cite{Koren}\cite{McCarthy} or score aggregation \cite{Baltrunas}\cite{Pizzutilo}.  
Preference aggregation aggregates the group users' preferences to make trade-off among group users' preferences.  
Existing score aggregate approaches are mainly \emph{average} \cite{Baltrunas}\cite{Berkovsky}, \emph{least misery} \cite{Roy}, \emph{maximum satisfaction} \cite{Boratto}. 
A major disadvantage of the memory-based approach is that it ignores the interaction among users; for example, the decision-making process in the group activity. 

In contrast, the model-based approaches \cite{Tran}\cite{Cao}\cite{YIN} utilize an attention mechanism to model the decision-making process. However, existing model-based approaches all assume the group is known and pre-defined, while real-life activity groups are usually temporary and form only for one or a few activities. \cite{Tran}. Furthermore, although the model-based group recommendations model the decision-making process, they ignore the impact of the target user (initiator) in a user group while the initiator's decision-making power can affect the final activity decision.

In summary, existing group recommendations are not only all single-typed, but also ignore the role of the initiator in the decision-making process of the group activities. 

\subsection{Geo-social Query}

The goal of geo-social query is to help the target user to find a user group satisfying social constraints. The spatial object indexing and query in spatio-temporal database have become the theoretical basis of geo-social queries in the geo-social network \cite{MZhang}\cite{XLiu}.
Most existing geo-social query approaches \cite{Sun}\cite{Yang} focus on selecting a user group from the target user's social connections while minimizing the distance to the gathering place; ensuring that selected users are in good social relationships or familiar with each other. Though geo-social query returns a user group with close social relationships, the query results are still single-typed.  

Recently, Yang et al. and Ma et al. expanded the single-typed geo-social query methods to multi-typed social group query\cite{Lee} \cite{Ma}. 
Yang et al. proposed the multi-typed social-temporal group query method, which returns the activity time combined with a user group under the social constraints that the group activity preference is satisfied\cite{Lee}. 
Ma et al.\cite{Ma} extended the single-typed $k$-cover geo-social group query proposed in \cite{Xu} to the multi-typed geo-social group query, whose query results are 2-tuples $(user group, location)$. 
Our proposed method differs from \cite{Ma} in the following aspects: 
a) geo-social group query lacks analysis on users' personalized features from individual check-in history data; b) the geo-social group query can not be applied to group activity planning since users' preferences are not taken into consideration.

\subsection{Event-Partner Recommendation}
Existing event-partner recommendation methods can be classified into three categories: given an event and the target user,  recommending a user who may join the event with the target user \cite{Liao}\cite{Tu}\cite{Ayala}; given the partners and the target user, recommending an event they may join together\cite{TU}; given the target user, recommending a two tuple \emph{(user, event)} \cite{Hongzhi}, i.e. finding a partner to participate in an event with the target user. 

The first two categories of event-partner both return single-typed recommendation results. Compared with the third category, our proposed method returns a group of user, thus can be applied to group activity planning, while event-partner recommendation method proposed in \cite{Hongzhi} returns only one user, which limits its application.

\section{Conclusion}

We propose AGSGR that recommends a user group along with top-$k$ activity locations at the same time. The user group is retrieved employing social constraints, and the activity locations are found based on the top-$k$ topics and the spatial accessibility of the recommended user group.
proposed method takes the target user's influence and geo-social acquaintance level into consideration, achieving dominating performance when compared with the baseline methods. The new type of recommendation may shed light on developing future applications on geo-social network and improving the existing location-aware services.

\end{document}